\documentclass[prd,twocolumn,preprint numbers, aps,nofootinbib]{revtex4}
 \usepackage{amsmath}
\usepackage{amssymb}
\usepackage{graphicx}
\usepackage{color}
\usepackage{bm}
\usepackage{times}
\usepackage{hyperref}
\hypersetup{
  colorlinks=true,
  citecolor=blue,
  linkcolor=blue,
  urlcolor=blue}

\usepackage{epsfig}
\usepackage{dcolumn}
\usepackage{float}

\usepackage{epsfig}
\usepackage{dcolumn}
\usepackage[normalem]{ulem}
\def\be{\begin{equation}}
\def\ee{\end{equation}}
\def\bea{\begin{eqnarray}}
\def\eea{\end{eqnarray}}
\def\gsim{\ \rlap{\raise 2pt\hbox{$>$}}{\lower 2pt \hbox{$\sim$}}\ }
\def\lsim{\ \rlap{\raise 2pt\hbox{$<$}}{\lower 2pt \hbox{$\sim$}}\ }
\def\dslash{\kern-4pt \not{\hbox{\kern-2pt $\partial$}}}
\def\pslash{\not{\hbox{\kern-2pt p}}}

\newcommand{\dcp}{\delta_{\rm CP}}

\newcommand{\cnv}{\v{C}erenkov}

\begin{document}

\DeclareGraphicsExtensions{.eps,.ps}

\title{A comparative study between ESSnuSB and T2HK in determining the leptonic CP phase}

\author{Monojit Ghosh}
\email{manojit@kth.se}
\affiliation{Department of Physics, School of Engineering Sciences,
KTH Royal Institute of Technology, AlbaNova University Center, Roslagstullsbacken 21, SE--106 91 Stockholm, Sweden}
\affiliation{The Oskar Klein Centre for Cosmoparticle Physics, AlbaNova University Center, Roslagstullsbacken 21, SE--106 91 Stockholm, Sweden}

\author{Tommy Ohlsson}
\email{tohlsson@kth.se}
\affiliation{Department of Physics, School of Engineering Sciences,
KTH Royal Institute of Technology, AlbaNova University Center, Roslagstullsbacken 21, SE--106 91 Stockholm, Sweden}
\affiliation{The Oskar Klein Centre for Cosmoparticle Physics, AlbaNova University Center, Roslagstullsbacken 21, SE--106 91 Stockholm, Sweden}

\begin{abstract}
In this paper, we perform a comparative analysis between the future proposed long-baseline experiments ESSnuSB and T2HK in measuring the leptonic CP phase $\delta_{\rm CP}$. In particular, we study the effect of the neutrino mass ordering degeneracy and the leptonic mixing angle $\theta_{23}$ octant degeneracy in the measurement of leptonic CP violation and precision for both experiments. Since the ESSnuSB (T2HK) experiment probes the second (first) oscillation maximum to study neutrino oscillations, the effect of these degeneracies are significantly different in both experiments. Our main conclusion is that for the ESSnuSB experiment, the information on the neutrino mass ordering does not play a major role in the determination of $\delta_{\rm CP}$, which is not the case for the T2HK experiment. However, the information on the true octant compromises the CP sensitivity of the ESSnuSB experiment as compared to T2HK if $\theta_{23}$ lies in the lower octant. These conclusions are true for both the 540~km and 360~km baseline options for the  ESSnuSB experiment. In addition, we investigate the effect of different running times in neutrino and antineutrino modes and the effect of $\theta_{23}$ precision in measuring $\delta_{\rm CP}$.
\end{abstract}

\maketitle

\section{Introduction}

Among the six parameters that describe the phenomenon of neutrino oscillations in the standard three-flavor framework, i.e., $\theta_{12}$, $\theta_{13}$, $\theta_{23}$, $\Delta m^2_{21}$, $\Delta m^2_{31}$, and $\dcp$, the remaining unknowns are the following: (i) the neutrino mass ordering, which can be either normal (NO, $\Delta m^2_{31} > 0$) or inverted (IO, $\Delta m^2_{31} < 0$), (ii) the octant of the leptonic mixing angle $\theta_{23}$, which can be either lower (LO, $\theta_{23}  < 45^\circ$) or higher (HO, $\theta_{23}  > 45^\circ$), and the leptonic Dirac CP-violating phase $\dcp$. The most recent global analysis of the world neutrino data shows a slight preference of NO over IO, whereas for the octant degeneracy all three possibilities are viable including $\theta_{23}=45^\circ$ \cite{Esteban:2018azc,Gariazzo:2018pei,Capozzi:2018ubv}. Regarding $\dcp$, the current best-fit value is around $217^\circ$ ($280^\circ$) for NO (IO) and the allowed 3$\sigma$ values lie between $135^\circ$ and $366^\circ$ ($196^\circ$ and $351^\circ$) for NO (IO) \cite{Esteban:2018azc}.

The ESSnuSB \cite{Baussan:2013zcy,Wildner:2015yaa} and the T2HK \cite{Abe:2016ero} are two future proposed neutrino long-baseline experiments, which are specifically designed to determine $\dcp$ with significantly large confidence level. Since both these experiments have baseline lengths less than 1000~km, their main goal will be solely focused on measuring $\dcp$. In the European ESSnuSB project, neutrinos will be produced with the power linac of the European Spallation Source (ESS) at Lund in Sweden. The two possible detector sites are the Garpenberg mine and the Zinkgruvan mine, which are located around 540~km and 360~km, respectively, from the neutrino source. On the other hand, the T2HK project will be an upgrade of the existing T2K experiment having a powerful beam and a large detector volume with a baseline of around 295~km. Both experiments will use megaton-class water-\cnv\ detectors. For T2HK, both the first oscillation maximum and the flux peak around 0.6~GeV, whereas for ESSnuSB, the flux peaks around 0.25~GeV, which is close to the second oscillation maximum for both baseline options. Since the variation of the neutrino oscillation probability with respect to $\dcp$ is much larger near the second oscillation maximum as compared to the first oscillation maximum, ESSnuSB has the advantage over T2HK to measure $\dcp$ with high precision despite of having lesser number of events. However, it is well known that the determination of $\dcp$ depends on the information on the neutrino mass ordering and the octant of $\theta_{23}$. The existence of an ordering-$\dcp$ degeneracy \cite{Prakash:2012az,Huber:2002mx} and an octant-$\dcp$ degeneracy \cite{Agarwalla:2013ju} can affect the determination of $\dcp$ if matter effects are not significant and the number of events is less. As both ESSnuSB and T2HK have baseline lengths less than 1000~km, they experience a lack of matter effects, and thus, it is not unexpected to believe that their leptonic CP violation measurement capability will be compromised. Furthermore, as ESSnuSB will have lesser number of events than T2HK, its capability can also suffer from the octant degeneracy. Note that these facts have been investigated in great detail for T2HK \cite{Agarwalla:2017nld,Fukasawa:2016yue,Raut:2017dbh,Chatterjee:2017xkb,Blennow:2014sja} and it has been shown that if the ordering is unknown, then the CP violation discovery is significantly less in T2HK for $0^\circ < \dcp < 180^\circ$ ($-180^\circ < \dcp < 0^\circ$) if the true ordering is NO (IO) \cite{Abe:2016ero}.\footnote{To solve this problem of T2HK, there is a proposal with the acronym T2HKK to place an additional detector in South Korea. This problem can also be solved by collecting atmospheric neutrino data in  the Hyper-Kamiokande detector \cite{Abe:2016ero}.} Moreover, there are a couple of studies regarding the $\dcp$ phase for ESSnuSB in the literature \cite{Agarwalla:2014tpa,Chakraborty:2017ccm}. However, to the best of our knowledge, a study of the effect of the ordering degeneracy and the octant degeneracy in measuring $\dcp$ for ESSnuSB has not been carried out yet.\footnote{For a study regarding the CP asymmetry at the second oscillation maximum, see Ref.~\cite{Bernabeu:2018use}.} This gives us the opportunity for a comparative study of parameter degeneracies at different oscillation maxima. In this paper, we have performed a detailed analysis of the CP sensitivity for both ESSnuSB and T2HK in terms of leptonic CP violation discovery and precision. Indeed, we have found that for ESSnuSB, the ordering-$\dcp$ degeneracy behaves in a different way. For the first time, we have shown that unlike for T2HK, for ESSnuSB the information on the neutrino mass ordering does not play much role in determining $\dcp$. However, the information on the octant of $\theta_{23}$ plays some role in determining $\dcp$, which is not the case for T2HK. We have also shown that these conclusions are true for both baseline options of ESSnuSB. Furthermore, we have studied how different neutrino running times affect the CP sensitivity for ESSnuSB and found that the dominant neutrino mode of ESSnuSB provides the best CP sensitivity. We have also studied the $\theta_{23}$  measurement capability of both experiments and showed how it affects the CP measurement for these experiments.  

This paper is organized in the following way. In Sec.~\ref{exp}, we will give the experimental and simulation details, which we will use in our analysis. Next, in Sec.~\ref{res}, we will discuss our main results. Finally, in Sec.~\ref{conc}, we will summarize and conclude.

\section{Experimental and Simulation Details}
\label{exp}

We use the software GLoBES \cite{Huber:2004ka} for the simulation of the ESSnuSB and T2HK experiments. For ESSnuSB, we use the same specification as given in Ref.~\cite{ess_deliverable}. We consider a proton beam of energy 2.5~GeV and power 5~MW capable of delivering $2.7 \times 10^{23}$ protons on target per year running for ten years. The far detector is a 507~kt water-\cnv\ tank located at either 540~km (Garpenberg mine) or 360~km (Zinkgruvan mine) away from the neutrino source (ESS). We also consider a near detector located at a distance of 500~m away from the neutrino source. We use correlated systematics between the near and far detectors. The details of the systematic errors are adopted from Ref.~\cite{Coloma:2012ji} and presented in Table~1.1 of Ref.~\cite{ess_deliverable}.
For T2HK, we consider a beam power of 1.3~MW with a total exposure of $27 \times 10^{21}$ protons on target, which corresponds to a ten-year running time. In this experiment, the neutrinos will be detected by two water-\cnv\ tanks with 187~kt mass each located a distance of 295~km away from the J-PARC source. Systematic errors are adopted from Table~VI of Ref.~\cite{Abe:2016ero}. 
Note that in our analysis we consider equal running time in neutrino and antineutrino modes (i.e., five years in neutrino mode and five years in antineutrino mode) for both experiments, unless otherwise stated. Our numerical calculations successfully match with the results of Refs.~\cite{Abe:2016ero,ess_deliverable}.

\section{Simulation Results}
\label{res}

Throughout this paper, we calculate the sensitivity in terms of a CP violation discovery $\chi^2$ function and a CP precision $\chi^2$ function. On one hand, CP violation discovery refers to the capability of an experiment to distinguish a particular value of $\dcp$ other than $0^\circ$ and $180^\circ$. On the other hand, CP precision refers to how well an experiment can separate between two different values of $\dcp$. The statistical $\chi^2$ function is defined as
\begin{equation}
 \chi^2_{{\rm stat}} = 2 \sum_{i=1}^n \bigg[ N^{{\rm test}}_i - N^{{\rm true}}_i - N^{{\rm true}}_i \log\bigg(\frac{N^{{\rm test}}_i}{N^{{\rm true}}_i}\bigg) \bigg]\,,
\end{equation}
where $n$ corresponds to the number of energy bins,  $N^{{\rm true}}$ is the number of true events, and $N^{{\rm test}}$ is the number of test events. We have incorporated the systematics by the method of pulls. In our analysis, the parameters $\theta_{12}$, $\theta_{13}$, and $\Delta m^2_{21}$ are kept fixed in both the true and test spectrum of the $\chi^2$ function. The values of these parameters are \cite{Esteban:2018azc,Gariazzo:2018pei,Capozzi:2018ubv}
 \begin{eqnarray} \nonumber
 \sin^2\theta_{12} &=& 0.312 \,,\\ \nonumber
 \sin^22\theta_{13} &=& 0.085 \,,\\ \nonumber
  \Delta m^2_{21} &=& 7.5 \times 10^{-5} {\rm eV}^2 \,. 
 \end{eqnarray}
The value of $\Delta m^2_{31} = 2.52 \times 10^{-3}$ eV$^2$ is fixed in the true spectrum and varied in its current $3 \sigma$ range in the test one. In the whole paper, we assume the true neutrino mass ordering to be normal, i.e., NO. For the case of `ordering known', the sign of $\Delta m^2_{31}$ is kept fixed in the test and for `ordering unknown', the sign is varied in the test. The canonical true values of $\theta_{23}$ are $42^\circ$, $45^\circ$, and $48^\circ$, corresponding to the LO, maximal mixing, and the HO, respectively. For `octant known', the value of $\theta_{23}$ is kept fixed (the value is the same for both test and true) and for `octant unknown', the test values of $\theta_{23}$ can have values lying in the correct octant as well as in the wrong octant, or in other words, the test values of $\theta_{23}$ are varied between $39^\circ$ and $51^\circ$.

\subsection{Results for $\mbox{\boldmath$\theta_{23}=45^\circ$}$}

\begin{figure*}
\hspace{-30pt}
\includegraphics[scale=1.1]{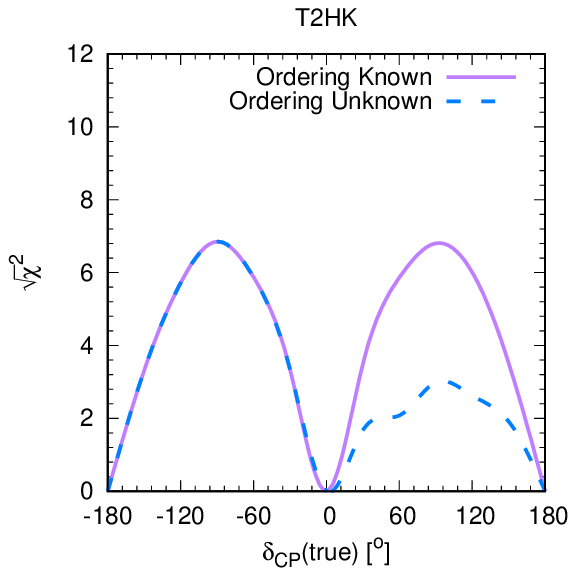}
\hspace{-80pt}
\includegraphics[scale=1.1]{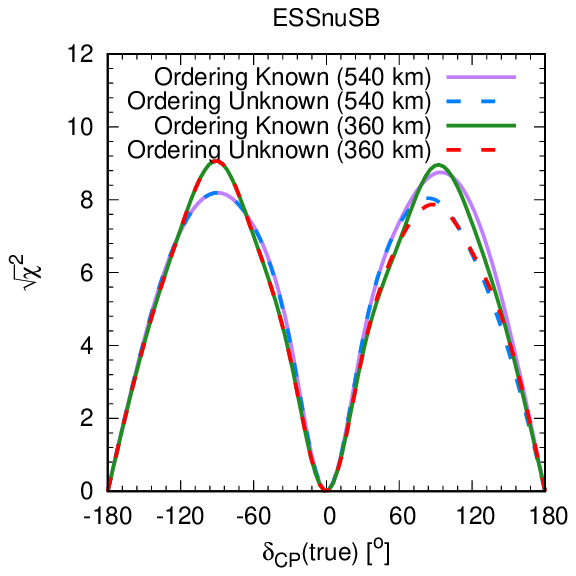}
\caption{CP violation sensitivity of ESSnuSB and T2HK for $\theta_{23}=45^\circ$ as a function of $\dcp\mbox{(true)}$. The left panel is for T2HK and the right panel is for ESSnuSB. The comparison between the sensitivities are shown when neutrino mass ordering is known versus when neutrino mass ordering is unknown. Note that the results are shown for normal neutrino mass ordering.}
\label{fig1}
\end{figure*}

\begin{figure*}
\hspace{-50pt}
\includegraphics[scale=0.91]{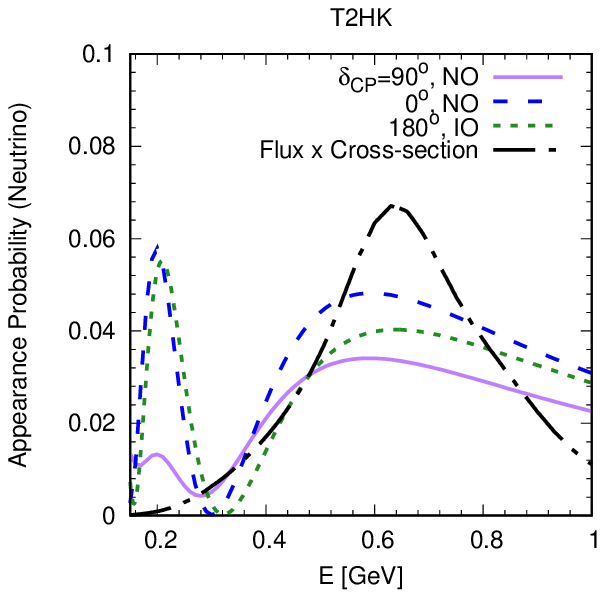}
\hspace{-70pt}
\includegraphics[scale=0.91]{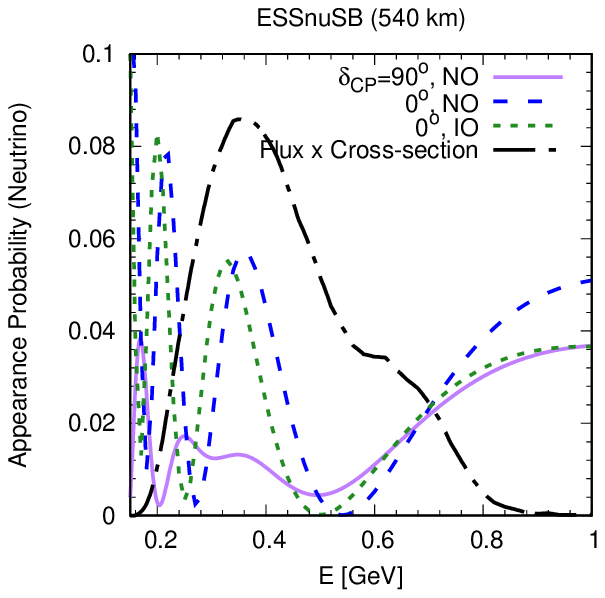}
\hspace{-70pt}
\includegraphics[scale=0.91]{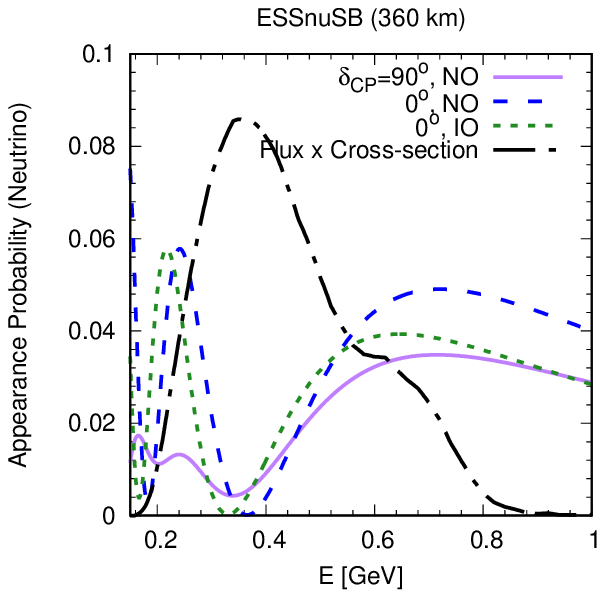}
\hspace{-50pt}
\caption{Appearance probability ($\nu_\mu \to \nu_e$) for $\theta_{23}=45^\circ$ as a function of neutrino energy $E$. The left, middle, and right panels are for T2HK, ESSnuSB (540~km), and ESSnuSB (360~km), respectively. All three panels are showing the neutrino appearance probability only. The black dash-dotted curves correspond to the product of the $\nu_\mu$ flux and the charged-current cross-section for $\nu_e$.}
\label{fig2}
\end{figure*}

First, let us discuss the case of maximal mixing, where there is no octant degeneracy. In this case, we have varied $\theta_{23}$ in its 3$\sigma$ range in the test. In the left (right) panel of Fig.~\ref{fig1}, we have plotted the CP violation discovery potential of T2HK (ESSnuSB) versus $\dcp\mbox{(true)}$. From the left panel, we observe that if the neutrino mass ordering is known, the CP violation discovery sensitivity is around $7 \sigma$ for both $\dcp = \pm 90^\circ$. However, if the neutrino mass ordering is unknown, the CP violation discovery sensitivity drops to $3 \sigma$ for $\dcp=90^\circ$. This is due to the well-known ordering-$\dcp$ degeneracy. However, this is evidently not the case for ESSnuSB. For ESSnuSB, the drop in the sensitivity around $\dcp=90^\circ$ is much less significant compared to T2HK when the neutrino mass ordering is unknown.
From the right panel, we note that for the ESSnuSB baseline option of 540~km (360~km), the CP violation discovery $\chi^2$ function falls from 75 (81) to 63 (62) if the neutrino mass ordering is unknown. 
Thus, we conclude that the information on the neutrino mass ordering does not play much role for the determination of $\dcp$ in ESSnuSB, and therefore, even if the ordering is unknown, it can discover CP violation with a significant confidence level in the unfavorable region of $\dcp$. 
This difference between ESSnuSB and T2HK arises due to the fact that the sensitivity of T2HK comes from the first oscillation maximum, whereas the sensitivity of the ESSnuSB comes from the second oscillation maximum. 
In this context, it is important to note that the fall of the sensitivity around $\dcp=90^\circ$ is slightly larger for the ESSnuSB baseline option of 360~km than for the one of 540~km. The reason for this can be understood from Fig.~\ref{fig2}.

\begin{figure*}
\hspace{-50pt}
\includegraphics[scale=0.91]{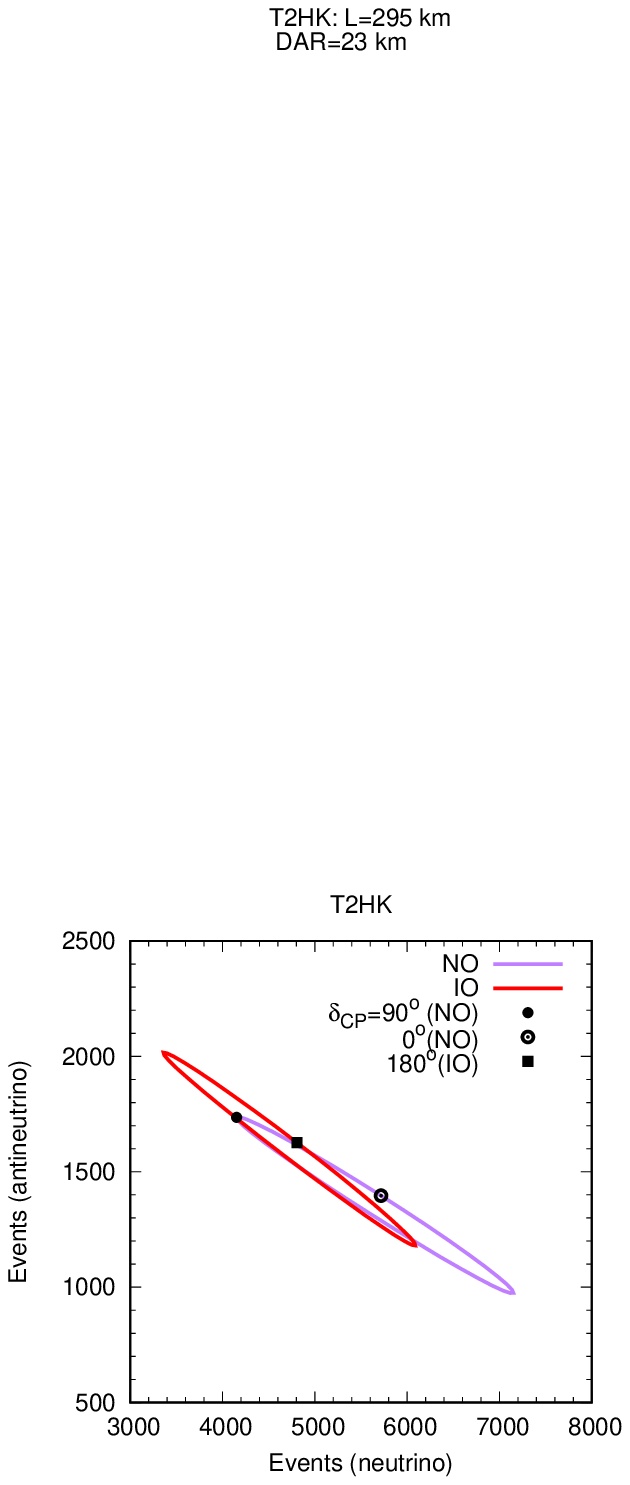}
\hspace{-70pt}
\includegraphics[scale=0.91]{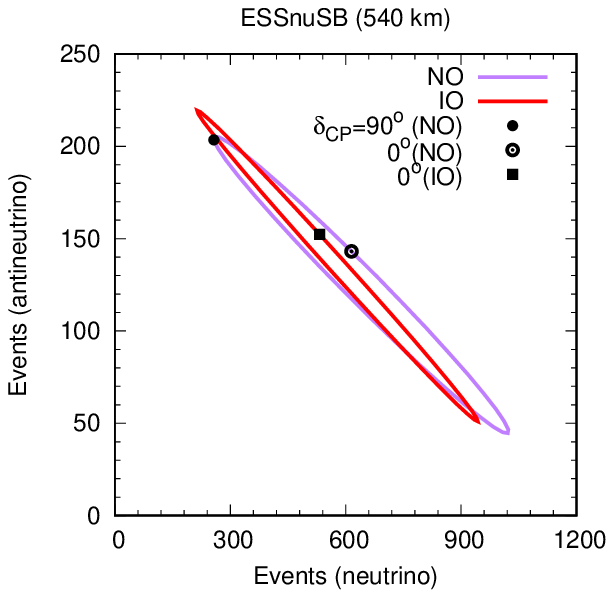}
\hspace{-70pt}
\includegraphics[scale=0.91]{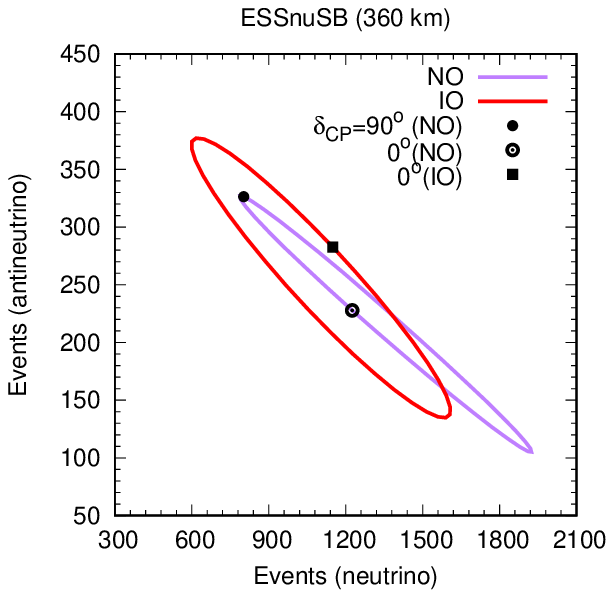}
\hspace{-50pt}
\caption{Bi-event plots for $\theta_{23}=45^\circ$. The left, middle, and right panels correspond to T2HK, ESSnuSB (540~km), and ESSnuSB (360~km), respectively. The purple ellipses are presented for normal neutrino mass ordering (NO) and the red ellipses are presented for inverted neutrino mass ordering (IO).}
\label{fig3}
\end{figure*}

\begin{figure*}
\hspace{-50pt}
\includegraphics[scale=0.91]{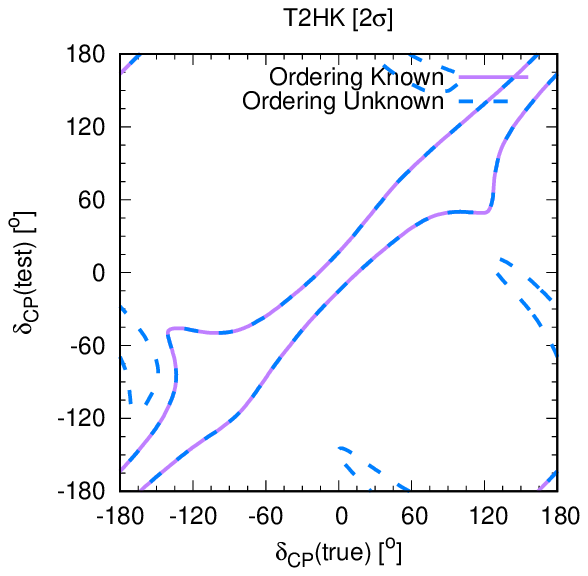}
\hspace{-70pt}
\includegraphics[scale=0.91]{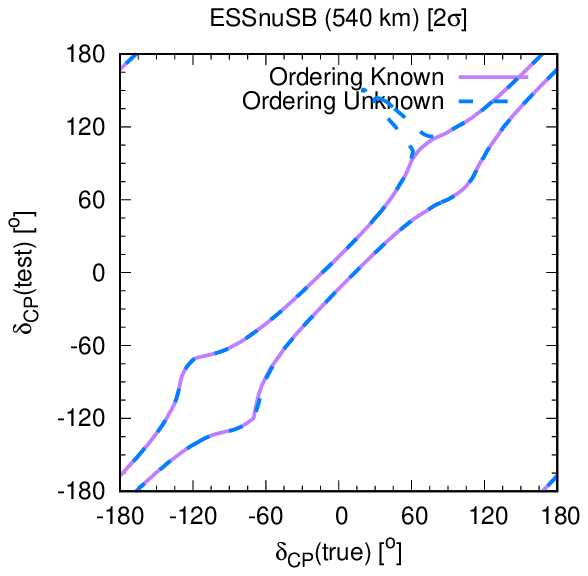}
\hspace{-70pt}
\includegraphics[scale=0.91]{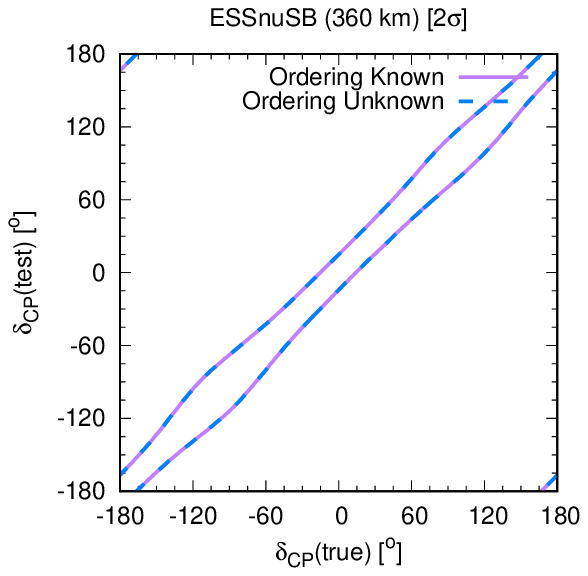}
\hspace{-50pt}
\caption{CP precision sensitivity of T2HK and ESSnuSB for $\theta_{23}=45^\circ$. The comparison in CP precision sensitivity is shown, when the neutrino mass ordering is known versus the neutrino mass ordering is unknown. All three panels are for normal neutrino mass ordering.}
\label{fig4}
\end{figure*}

In Fig.~\ref{fig2}, we have plotted the appearance channel probability ($\nu_\mu \rightarrow \nu_e$) as a function of energy. Since the physics of the ordering-$\dcp$ degeneracy is same for both neutrinos and antineutrinos, we only present the plots for neutrinos. The left panel is for T2HK, whereas the middle and right panels are for the ESSnuSB baseline options of 540~km and 360~km, respectively. In all three panels, we have given the product of the $\nu_\mu$ flux and the charged-current cross-section for $\nu_e$ (hereafter denoted `flux $\times$ cross-section') in arbitrary units to depict energies which are relevant for our discussion. The different curves in the panels are displayed to specifically understand the results of Fig.~\ref{fig1}. From Fig.~\ref{fig1}, we have seen that for NO, the sensitivity is affected for $\dcp\mbox{(true)}$ value of $90^\circ$. Therefore, in all three panels, the curve corresponding to $\dcp = 90^\circ$ is the true reference curve. For CP violation discovery, the $\chi^2$ minimum appears at either $\dcp\mbox{(test)}$ values of $0^\circ$ or $180^\circ$. We have checked that for ESSnuSB, the $\chi^2$ minimum always appears when $\dcp\mbox{(test)}$ equals $0^\circ$ for both cases, i.e., ordering is known and ordering is unknown. For T2HK, if the ordering is known, the  $\chi^2$ minimum appears for $\dcp\mbox{(test)} = 0^\circ$, whereas if ordering is unknown, the $\chi^2$ minimum appears for $\dcp\mbox{(test)} = 180^\circ$. Now, note that `ordering known' implies the fact that the $\chi^2$ minimum comes in the true ordering, which is NO in our case, and `ordering unknown' implies that the $\chi^2$ minimum comes in the wrong ordering, which is IO in our case. Therefore for test reference points, we have plotted the curves for ($0^\circ$, NO) and ($0^\circ$, IO) for ESSnuSB (middle and right panels) and ($0^\circ$, NO) and ($180^\circ$, IO) for T2HK (left panel). The separation between the $\dcp = 90^\circ$ curve and the curve for NO corresponds to the sensitivity if the ordering is known and the separation between the $\dcp = 90^\circ$ curve and the curve for IO corresponds to the sensitivity when the ordering is unknown. The first thing, which we observe from the panels, is that for T2HK, the flux $\times$ cross-section peaks at the first oscillation maximum, whereas for ESSnuSB, the flux $\times$ cross-section peaks at the second oscillation maximum for the baseline option of 540~km and covers some part of the first and second oscillation maxima for the baseline option of 360~km.\footnote{Note that although the ESSnuSB flux peaks around 0.25~GeV, the charged-current cross-section is almost negligible at this energy. Therefore, the effective energy distribution, where the number of events is maximal, appears in the region where the flux $\times$ cross-section peaks, which is around 0.35~GeV \cite{Wildner:2015yaa}.} Furthermore, we see that for T2HK, the ($180^\circ$, IO) curve is much closer to the $\dcp = 90^\circ$ curve than to the ($0^\circ$, NO) curve around the first oscillation maximum, where the flux peaks. Thus, if the ordering is unknown, the sensitivity falls drastically for T2HK around $\dcp = 90^\circ$. However, this is not the case for ESSnuSB. For the ESSnuSB baseline option of 540~km, 
the flux peaks around the second oscillation maximum, where the ($0^\circ$, NO)  and ($0^\circ$, IO) curves are almost overlapping. Therefore, the separation between $\dcp = 90^\circ$ for the ($0^\circ$, NO) and ($0^\circ$, IO) curves are almost the same and very large. 
For the ESSnuSB baseline option of 360~km, the sensitivity comes from both the first and second oscillation maxima and the separation of the ($0^\circ$, NO) and ($0^\circ$, IO) curves from the $\dcp = 90^\circ$ curve is large for most of the region, where the flux $\times$ cross-section peaks.
Due to this reason, the sensitivity of ESSnuSB is not so much reduced if the ordering is unknown, and therefore, it is capable of measuring leptonic CP violation with high precision without any information on the neutrino mass ordering. 
Note that unlike the ESSnuSB baseline option of 540~km, for the one of 360~km, there is an overlap between the ($0^\circ$, NO) and ($0^\circ$, IO) curves with the $\dcp = 90^\circ$ curve in some of the region and that is why the fall in the sensitivity for the option of 360~km is larger than for the option of 540~km if the neutrino mass ordering is unknown.
It is also important to understand that for both the baseline options of ESSnuSB, the ($0^\circ$, IO) curve is much closer to the $\dcp = 90^\circ$ curve than to the ($0^\circ$, NO) curve near the first oscillation maximum. Therefore, if ESSnuSB had a flux that peaks near the first oscillation maximum, this advantage could have been lost and the result would have been similar to that of T2HK.

\begin{figure*}
\includegraphics[scale=1.1]{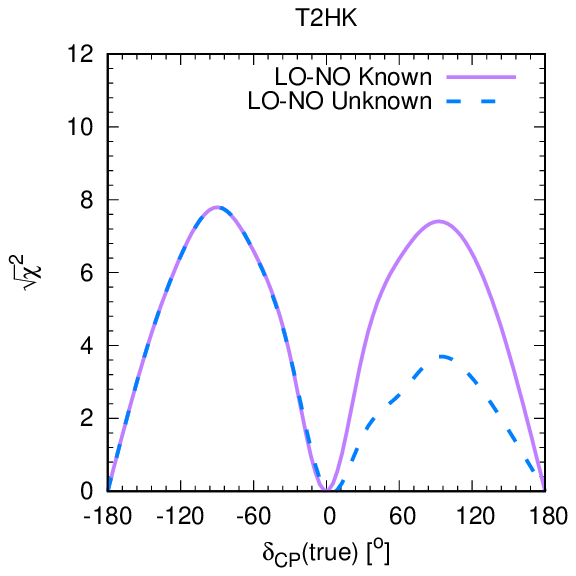}
\hspace{-80pt}
\includegraphics[scale=1.1]{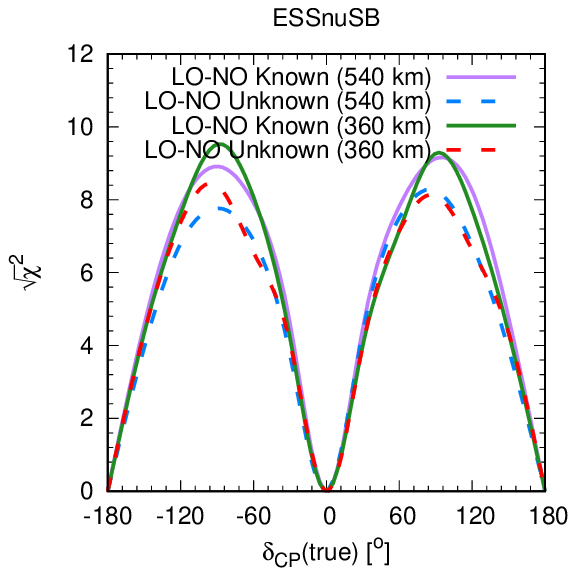}
\includegraphics[scale=1.1]{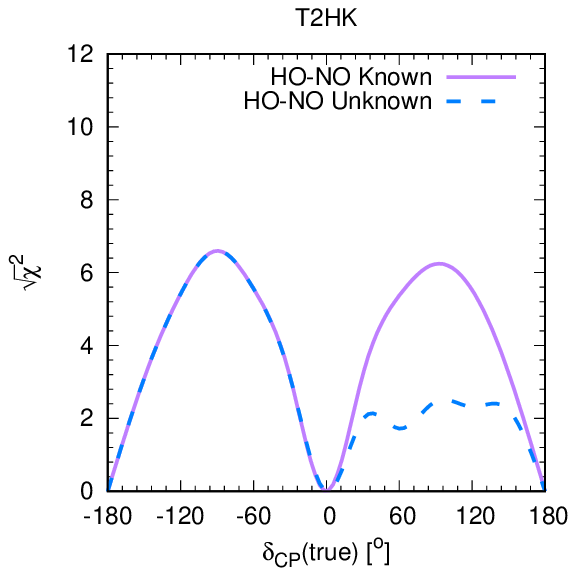}
\hspace{-80pt}
\includegraphics[scale=1.1]{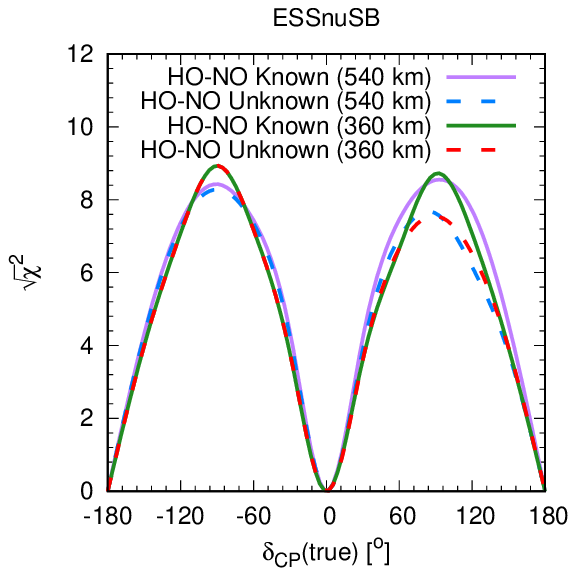}
\caption{CP violation sensitivity of ESSnuSB and T2HK for $\theta_{23} \in \{42^\circ,48^\circ\}$ as a function of $\dcp\mbox{(true)}$ if ordering and octant are both unknown. All panels are for normal neutrino mass ordering (NO). The upper-left (lower-left) panel is for T2HK if the true octant is the lower (higher) octant LO (HO), whereas the upper-right (lower-right) panel is for ESSnuSB if the true octant is the lower (higher) octant LO (HO). }
\label{fig5}
\end{figure*}

Another simple way to understand the above fact is that for ESSnuSB the oscillation probabilities for NO and IO are almost identical at the second oscillation maximum. We know that matter effects are proportional to neutrino energy and the second oscillation maximum for ESSnuSB occurs at relatively lower energies than the ones at the first oscillation maximum. For this reason at the second oscillation maximum, matter effects are too weak to cause any separation between the two orderings. This fact will be even more evident from Fig.~\ref{fig3}, where we have presented the bi-event plots. These plots will also depict the role of antineutrinos and event samples involved in both experiments. On the x-axis (y-axis), we have plotted the appearance channel events for neutrinos (antineutrinos). The trajectory of $\dcp$ in this plane produces an elliptical shape \cite{Minakata:2001qm}. In the left panel, we have given the curves of T2HK, whereas in the middle and right panels, we have given the curves of ESSnuSB for the baseline options of 540~km and 360~km, respectively. In each panel, the purple and red ellipses correspond to NO and IO, respectively. Similarly to Fig.~\ref{fig2}, the ($90^\circ$, NO), ($0^\circ$, NO), and ($180^\circ$, IO) points are marked for T2HK and the ($90^\circ$, NO), ($0^\circ$, NO), ($0^\circ$, IO) points are marked for ESSnuSB. Immediately, we observe from the panels that although the number of events for ESSnuSB is significantly less than for T2HK, still the CP sensitivity of ESSnuSB is much higher than that of T2HK. This is due to the variation of the probability with respect to $\dcp$ which is much larger at the second oscillation maximum than at the first oscillation maximum, which can be also understood by looking at the bi-event plots. 
In Fig.~\ref{fig3}, for T2HK (left panel), we note that the point ($90^\circ$, NO) is much closer to ($180^\circ$, IO) than to ($0^\circ$, NO). For this reason, if the neutrino mass ordering is unknown, the CP sensitivity at $\dcp=90^\circ$ is drastically reduced. For the ESSnuSB baseline option of 540~km (middle panel), we observe that the ellipses for NO and IO are almost identical, and therefore, the separations of ($90^\circ$, NO) from ($0^\circ$, NO) and ($0^\circ$, IO) are not so different. Thus, the CP sensitivity does not depend on the fact that the neutrino mass ordering is known or unknown. On the other hand, for the ESSnuSB baseline option of 360~km (right panel), the separations of ($90^\circ$, NO) from ($0^\circ$, NO) and ($0^\circ$, IO) are longer than for the ESSnuSB baseline option of 540~km, but shorter than for T2HK. Therefore, the fall in the sensitivity if the neutrino mass ordering is unknown is larger for the ESSnuSB baseline option of 360~km than for the one of 540~km, but still the reduction in the sensitivity is not as severe as for T2HK.

\begin{figure*}
\hspace{-50pt}
\includegraphics[scale=0.91]{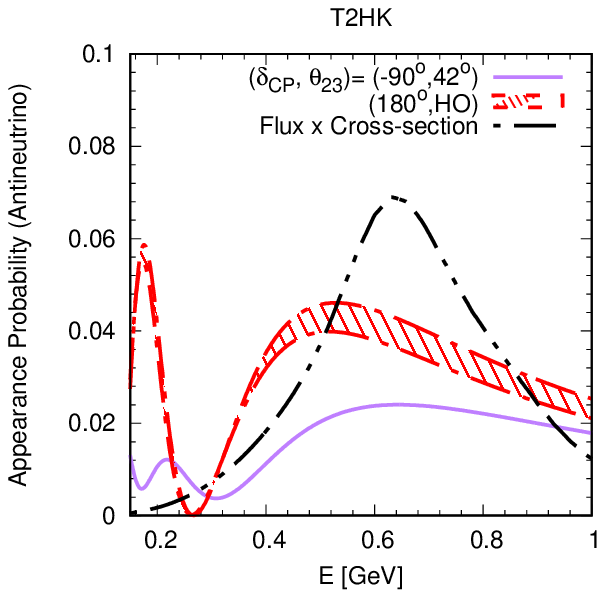}
\hspace{-70pt}
\includegraphics[scale=0.91]{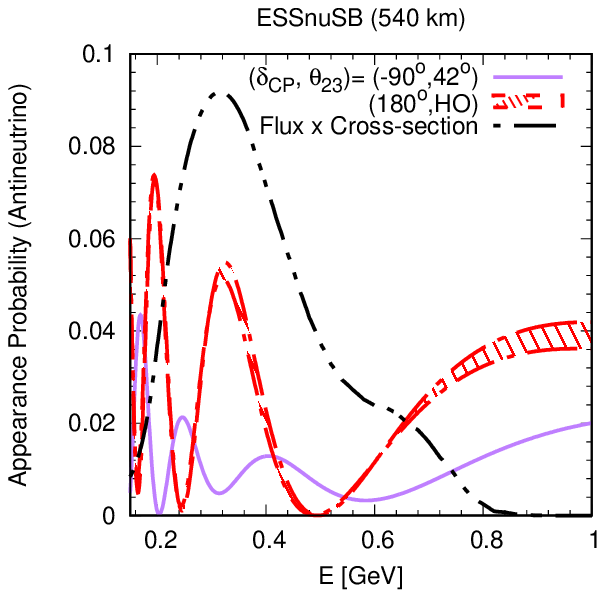}
\hspace{-70pt}
\includegraphics[scale=0.91]{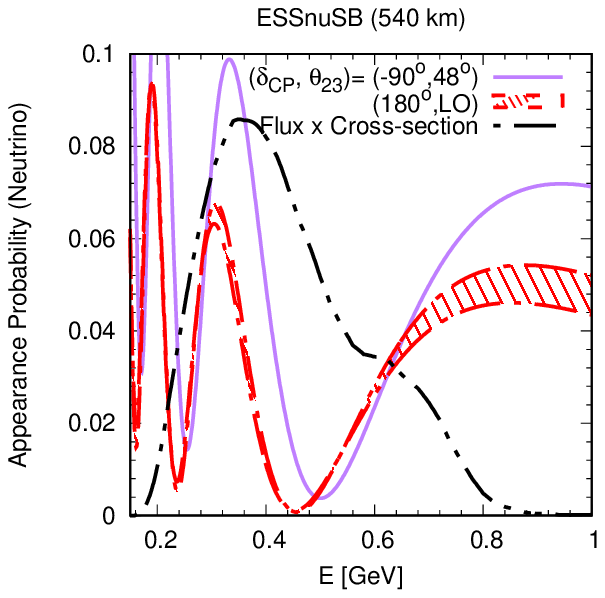}
\hspace{-50pt}
\caption{Appearance probability for $\theta_{23} \neq 45^\circ$ as a function of neutrino energy $E$ for T2HK and the ESSnuSB baseline option of 540~km. The left and middle panels correspond to the antineutrino probability. The purple curve is presented for the LO and the red band is presented for the HO. The right panel corresponds to the neutrino probability. The purple curve is presented for the HO and the red band is presented for the LO. In all panels, flux $\times$ cross-section is shown by black dash-dotted curves.}
\label{fig6}
\end{figure*}

Finally, in Fig.~\ref{fig4}, we have plotted the 2$\sigma$ CP precision contours in the $\dcp\mbox{(true})$ versus $\dcp\mbox{(test)}$ plane.\footnote{ We have checked that at 1$\sigma$ degenerate solutions do not appear in the CP precision for T2HK. Since degenerate solutions in the CP precision for T2HK start to appear from 2$\sigma$, we have chosen to present the CP precision contours at 2$\sigma$.} The left panel is for T2HK and the middle (right) panel is for the ESSnuSB baseline option of 540~km (360~km). In each panel, the purple and blue curves correspond to the cases of ordering is known and ordering is unknown, respectively. The panels can be understood in the following way. Each point on the x-axis is the true value and its corresponding width along the y-axis is the 2$\sigma$ uncertainty associated with this value. In the ideal situation, the allowed points should be located along the diagonal, where $\dcp\mbox{(true)} = \dcp\mbox{(test)}$. The common feature in all three panels is that the CP precision is best around $\dcp\mbox{(true)} = 0^\circ$ and worst around $\dcp\mbox{(true)} = \pm 90^\circ$. From the left panel, we see that if ordering is unknown, the CP precision capability of T2HK gets deteriorated significantly compared to if ordering is known. There are off-diagonal spurious regions that are allowed if the information on the ordering is absent. This is the effect of the ordering-$\dcp$ degeneracy, which is present at the first oscillation maximum. However, as we discussed earlier, at the second oscillation maximum, the information on the ordering does not affect the CP measurement, and from the right panel, we see that for the ESSnuSB baseline option of 360~km, the contours for `ordering known' and `ordering unknown' are exactly identical. However, from the middle panel, we see that for the ESSnuSB baseline option of 540~km, there is a small wiggle around $90^\circ$ if ordering is unknown. Therefore, we can also say that for ESSnuSB it possible to measure CP precision at a very high confidence level, even without the information on the ordering. Note that the CP precision (also CP violation) capability of the ESSnuSB baseline option of 360~km is slightly better than the ESSnuSB baseline option of 540~km. This is due to the fact that for the 360~km baseline, the number of events is larger by factor of $(540/360)^2 \sim 2$ as compared to the 540~km baseline. 

\subsection{Results for $\mbox{\boldmath$\theta_{23} \neq 45^\circ$}$}

In this section, we will study how the information on the octant affects the CP sensitivity capability of ESSnuSB and T2HK. In Fig.~\ref{fig5}, we have presented the CP violation discovery potential of ESSnuSB and T2HK if $\theta_{23} \neq 45^\circ$. The upper-row panels are for the LO and the lower-row panels are for the HO. In each row, the left panel is for T2HK and the right panel is for ESSnuSB. In these panels, we have compared the sensitivities if both ordering and octant are known versus if both ordering and octant are unknown. First, let us discuss the case of T2HK. From the left-column panels, we see that for the unknown case, there is a drop in the sensitivity around $\dcp=90^\circ$, but there is no reduction in the sensitivity for $\dcp=-90^\circ$. From the earlier discussion, we understand that the drop in the sensitivity around $\dcp=90^\circ$ is due to the ordering degeneracy. Therefore, we conclude that for T2HK the information on the octant does not play any role in the discovery of CP violation. This is true for both the LO and the HO. However, this is not the case for ESSnuSB. From the right-column panels, we see that except for the small drop in the sensitivity around $\dcp=90^\circ$, there is also a reduction in the sensitivity around $\dcp=-90^\circ$ if the true octant is the LO. For the ESSnuSB baseline option of 540~km (360~km), the sensitivity falls from 
9$\sigma$ (9.5$\sigma$) to 7.5$\sigma$ (8$\sigma$). However for the HO, the CP sensitivity of ESSnuSB does not depend on the information on the octant for both baseline options. 

Now, we will try to understand these features from the probability plots. In Fig.~\ref{fig6}, we have plotted the appearance channel probability as a function of energy for T2HK and the ESSnuSB baseline option of 540~km. Since our aim is to understand the effect of the octant degeneracy, all curves in the panels are presented for NO. From Fig.~\ref{fig5}, we understood that the effect of the octant degeneracy affects the CP measurement capability around $\dcp=-90^\circ$. Therefore, in Fig.~\ref{fig6}, the $(\dcp,\theta_{23}) = (-90^\circ, 42^\circ)$ and $(\dcp,\theta_{23}) = (-90^\circ, 48^\circ)$ curves represent the true points for the LO and the HO, respectively. From our understanding of the parameter degeneracy \cite{Barger:2001yr,Ghosh:2015ena}, we know that for ($-90^\circ$, $42^\circ$) there is an octant degeneracy in the neutrino oscillation probability, while the antineutrino oscillation probability is free from such an octant degeneracy. This implies that for the LO antineutrinos are responsible for the removal of degeneracies. Therefore, in the left panel, we have given the antineutrino oscillation probability for T2HK in order to understand how the octant degeneracy is resolved in the particular case of the LO. Since we have checked that the $\chi^2$ minimum occurs for $\dcp=180^\circ$, we have plotted the ($180^\circ$, HO) band corresponding to the test point. The HO band is due to the variation of $\theta_{23}$ from $45^\circ$ to $50^\circ$. The separation between the purple curve and the red band in the energy region where the flux peaks corresponds to the octant sensitivity. From the left panel, we see that there is a clear separation between the purple curve and the red band around 0.6 GeV and this is the reason why for T2HK the information on the octant does not play any role in the determination of CP violation discovery. However, for ESSnuSB, we see that for the LO (middle panel), there is an overlap between the purple curve and red band in the region where the flux $\times$ cross-section peaks. This is why the CP violation discovery is getting compromised around $\dcp=-90^\circ$ for the LO if the octant is unknown. However, let us understand why this is not happening for the HO. Again, from the knowledge on the octant degeneracy, we know that for  ($-90^\circ$, $48^\circ$) there is an octant degeneracy in the antineutrino oscillation probability and the neutrino oscillation probability is free from such a degeneracy. This implies that for the HO neutrinos are responsible for the removal of degeneracies. Therefore, in the right panel, we have plotted the neutrino oscillation probability to understand the effect for the HO. The red band corresponds to ($180^\circ$, LO). The LO band is due to the variation of $\theta_{23}$ from $40^\circ$ to $45^\circ$. We have also checked that the $\chi^2$ minimum occurs at $\dcp=180^\circ$. In addition, we see that there is an overlap between the purple curve and red band in some of the region where the flux $\times$ cross-section peaks. However, it is important to note that the values of the probability in the right panel are larger than the values of the probability in the middle panel. Also, since the neutrino cross-section is almost three times larger than the antineutrino cross-section, the neutrino run is sufficient to lift the degeneracy for the HO. Therefore, for the LO, the degeneracy is lifted by antineutrinos, whereas for the HO, the degeneracy is lifted by neutrinos. Henceforth the lack of events in the antineutrino run is insufficient to lift the degeneracy for the LO, and therefore, the CP sensitivity for ESSnuSB is affected by information on the octant for the LO. This does not happen for T2HK, since the number of antineutrino events is high because of (i) a clear removal of the octant degeneracy at the first oscillation maximum and (ii) large statistics. 

\begin{figure}
\includegraphics[scale=1.1]{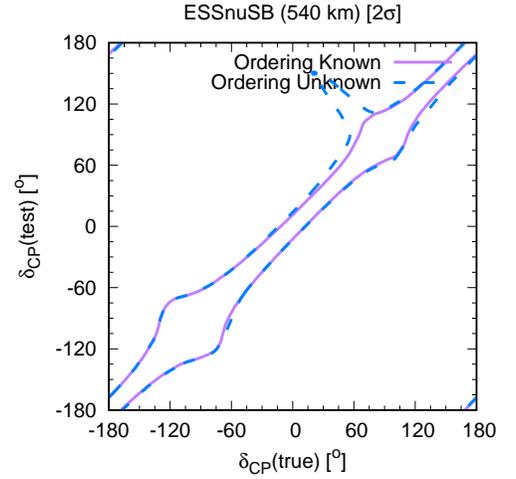}
\caption{CP precision plot for the ESSnuSB baseline option of 540~km in the plane of $\dcp\mbox{(true)}$ versus $\dcp\mbox{(test)}$. The true value of $\theta_{23}$ is $42^\circ$, which lies in the lower octant. The purple (blue) $2\sigma$ contour corresponds to the case if both ordering and octant are known (unknown).}
\label{fig7}
\end{figure}

Finally, let us see how the information on the octant affects the CP precision measurement capability of ESSnuSB. In Fig.~\ref{fig7}, we have only presented the case of the LO for the baseline option of 540~km. The purple (blue) 2$\sigma$ contour corresponds to the case if both ordering and octant are known (unknown). From this figure, we clearly see that at $2\sigma$ confidence level, the CP precision does not alter much, depending on the information on ordering and octant.  Since this figure is similar to that of the middle panel in Fig.~\ref{fig4}, we can definitely say that at $2 \sigma$ confidence level, the information on the octant does not affect the CP precision of ESSnuSB even for the LO.

\begin{figure}
\includegraphics[scale=1.1]{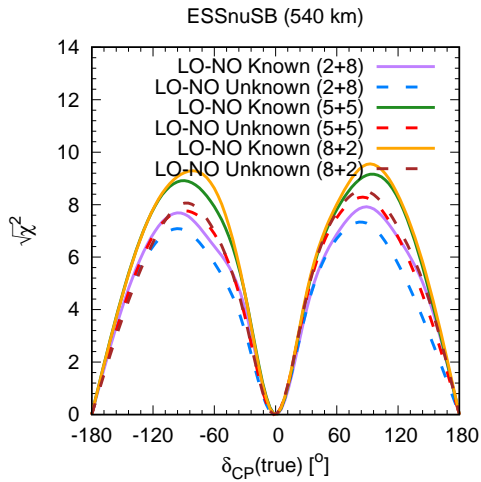}
\caption{CP violation sensitivity for the ESSnuSB baseline option of 540~km as a function of $\dcp\mbox{(true)}$. The true value of $\theta_{23}$ is $42^\circ$, which lies in the lower octant. Different running times in neutrino and antineutrino modes are considered. `a+b' corresponds to `a' years running time in neutrino mode and `b' years running in antineutrino mode.}
\label{fig8}
\end{figure}

\begin{figure*}
\hspace{-30pt}
\includegraphics[scale=1.1]{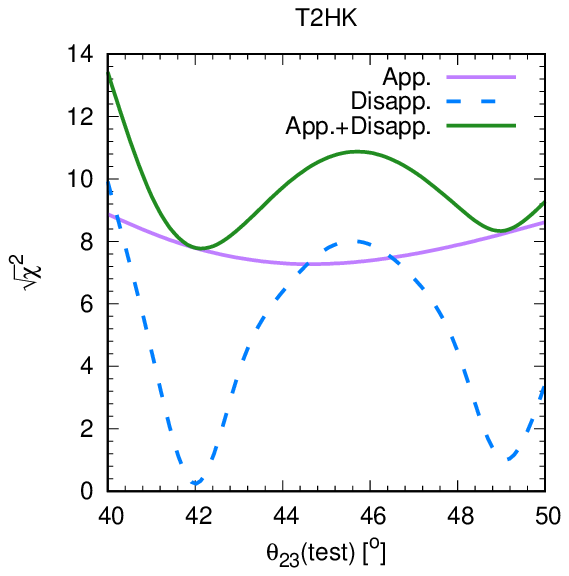}
\hspace{-80pt}
\includegraphics[scale=1.1]{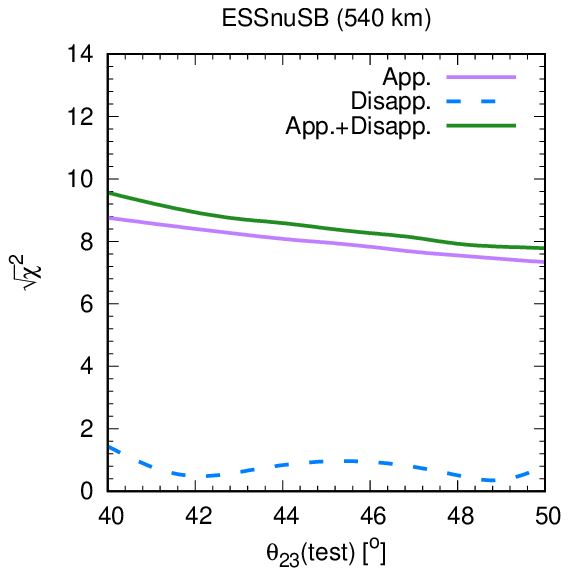}
\caption{CP violation sensitivity as a function of $\theta_{23}\mbox{(test)}$ for T2HK and the ESSnuSB baseline option of 540~km. The curves are presented for $\dcp\mbox{(true)} = -90^\circ$ and $\theta_{23}\mbox{(true)} = 42^\circ$.}
\label{fig9}
\end{figure*}

\subsection{Effect of different neutrino running times for the lower octant}

In the previous section, we have seen that for ESSnuSB, the CP violation discovery sensitivity around $\dcp=-90^\circ$ is compromised for the LO and we understood that this is because of the insufficient antineutrino events. In this section, we will study what happens if ESSnuSB runs in a dominant neutrino or antineutrino mode. We will study this for $\theta_{23}=42^\circ$ and the ESSnuSB baseline option of 540~km. 

In Fig.~\ref{fig8}, we have plotted the CP violation discovery $\chi^2$ function for three different running options for ESSnuSB, which are 2+8 (dominant antineutrino mode), 5+5 (equal neutrino and antineutrino modes), and 8+2 (dominant neutrino mode). Here, `a+b' corresponds to `a' years running time in neutrino mode and `b' years running time in antineutrino mode. In this figure, the solid curves correspond to both ordering and octant known, whereas the dashed curves correspond to both ordering and octant unknown. For our discussion, we will focus around the $\dcp=-90^\circ$ region. We clearly observe that for the dominant antineutrino mode (2+8), the drop in the sensitivity if LO-NO is unknown as compared to if LO-NO is known is the smallest than for the other two cases of 5+5 and 8+2. This confirms our argument that for the LO, antineutrinos are necessary to remove the octant degeneracy. However, due to the dominant antineutrino mode, there is a reduction of overall statistics, and therefore, this running time option of ESSnuSB provides the lowest sensitivity. On the other hand, for the dominant neutrino mode (8+2), although the drop in the sensitivity is the largest if LO-NO is unknown as compared to if LO-NO is known but due to higher statistics, the sensitivity is also the highest if LO-NO is unknown. For equal neutrino and antineutrino modes (5+5), the sensitivity is better than 2+8 but worse than 8+2. Therefore, from this discussion, we can conclude that although the dominant antineutrino mode is necessary to remove the octant degeneracy, the dominant neutrino mode provides the best sensitivity of ESSnuSB. 

\subsection{Effect of $\mbox{\boldmath$\theta_{23}$}$ precision for the lower octant}

In this section, we will study the capability of ESSnuSB and T2HK to measure $\theta_{23}$ and its  effect on the CP violation discovery sensitivity. We will analyse this for $\dcp\mbox{(true)} = -90^\circ$ and $\theta_{23}\mbox{(true)} = 42^\circ$. We will assume equal running time in neutrino and antineutrino modes (5+5) for both experiments.

In Fig.~\ref{fig9}, we have plotted the CP violation discovery $\chi^2$ function as a function of $\theta_{23}\mbox{(test)}$. The left panel is for T2HK, whereas the right panel is for the ESSnuSB baseline option of 540 km. In the two panels, we have displayed the appearance channel ($\nu_\mu \to \nu_e$) $\chi^2$ function, the disappearance channel ($\nu_\mu \to \nu_\mu$) $\chi^2$ function, and the combined appearance and disappearance ($\nu_\mu \to \nu_e$  and $\nu_\mu \to \nu_\mu$) $\chi^2$ function. The true ordering is assumed to be unknown. The global minimum of the combined appearance and disappearance curve is the CP violation discovery sensitivity corresponding to $\dcp\mbox{(true)} = -90^\circ$ and $\theta_{23}\mbox{(true)} = 42^\circ$ if both ordering and octant are unknown. Now, it is well known that the octant and CP sensitivity comes mainly from the appearance channel and the precision of $\theta_{23}$ stems from the disappearance channel. From the panels, we observe that for T2HK, the disappearance channel $\chi^2$ function is very sharp around $\theta_{23}\mbox{(test)} = 42^\circ$ and $49^\circ$. This provides a good measurement of $\theta_{23}$. On the other hand, as mentioned earlier, the appearance channel provides a good CP sensitivity. Note that the appearance channel $\chi^2$ function increases for $\theta_{23}\mbox{(test)} > 45^\circ$. This is due to the removal of the octant degeneracy by the antineutrino run. Then, if both channels are combined, because of the steep nature of the disappearance channel $\chi^2$ function, there appears two local minima and the global minimum materializes at the correct value of $\theta_{23}$. However, for ESSnuSB, the disappearance channel $\chi^2$ function is very shallow, and therefore unlike the one for T2HK, it does not provide a good measurement of $\theta_{23}$. On the other hand, the appearance channel provides a good CP sensitivity, but this is a decreasing function of $\theta_{23}\mbox{(test)}$. This is due to the fact that the antineutrinos are not sufficient to lift the octant degeneracy. However, note that because of the shallow nature of the disappearance channel $\chi^2$ function, the combined $\chi^2$ function does not have any local minimum and the $\chi^2$ minimum occurs at $\theta_{23}\mbox{(test)} = 50^\circ$. Therefore, if the disappearance channel $\chi^2$ function had a steeper nature like the one for T2HK, it would have caused two $\chi^2$ minima around $42^\circ$ and  $49^\circ$, resulting in an enhancement of the CP sensitivity. Thus, from this discussion, we understand that the $\theta_{23}$ precision measurement capability of ESSnuSB is inferior to that of T2HK and this also affects the CP sensitivity of ESSnuSB. One way to improve the $\theta_{23}$ measurement capability of ESSnuSB is by adding the atmospheric neutrino data sample~\cite{Chakraborty:2019jlv}.

\section{Summary and Conclusions}
\label{conc}

In this paper, we have performed a thorough comparative study between the ESSnuSB and T2HK experiments to measure the leptonic CP phase $\dcp$. In particular for the first time, we have studied the effect of both the ordering and octant degeneracies on the CP measurement capability of these two experiments. We have presented our results in terms of both CP violation and CP precision. It is well known that the capability of T2HK to discover CP violation is limited due to its lack of information on the neutrino mass ordering. However, our analysis show that for ESSnuSB the CP measurement capability is almost independent of the ordering. This is mainly due to the fact that T2HK is designed to study neutrino oscillations at the first oscillation maximum, whereas ESSnuSB is designed to study neutrino oscillations at second oscillation maximum. At the first oscillation maximum, the separation between the CP conserving phases for inverted ordering and the CP violating phase $\dcp=90^\circ$ for normal ordering is very small, but this separation is quite high near the second oscillation maximum. We have found that this is true for both baseline length options of ESSnuSB, which are 540~km and 360~km. Regarding the effect of the octant degeneracy, we have found that the CP sensitivity of T2HK is not affected by the information on the octant. However, our study shows that for ESSnuSB the CP sensitivity around $\dcp=-90^\circ$ is compromised for the lower octant as compared to T2HK if the octant is unknown. For the higher octant, the information on the octant does not play any role. This is due to the fact that for the lower octant, the octant degeneracy is resolved by antineutrinos at the first oscillation maximum, whereas at the second oscillation maximum, there is some overlap between the probability for $\dcp=-90^\circ$  and the higher octant. The small number of antineutrino events of ESSnuSB for the lower octant is not capable of removing the octant degeneracy at the second maximum and this is why the sensitivity is reduced if the octant is unknown. However, this degeneracy does not affect the CP precision capability of ESSnuSB at $2\sigma$ confidence level. To understand this point further, we have analysed the sensitivity of the ESSnuSB baseline option of 540~km for different running times in neutrino and antineutrino modes for the lower octant. We have found that for the dominant antineutrino mode, i.e., two years running time in antineutrino mode and eight years running time in neutrino mode (2+8), the drop in the CP sensitivity around $\dcp=-90^\circ$ is minimal as compared to other options if both ordering and octant are unknown. However, due to lack of overall statistics, 2+8 provides the worst sensitivity among the other options. The best sensitivity of the ESSnuSB baseline option of 540~km comes from the dominant neutrino mode, i.e., 8+2. Furthermore, we have shown that the $\theta_{23}$ precision capability of the ESSnuSB baseline option of 540~km is inferior to that of T2HK and it also affects the CP sensitivity of ESSnuSB. The results and analysis presented in this work help to understand the behavior of the ordering and octant degeneracies around the first and second oscillation maxima.


\begin{acknowledgments}
The authors would like to thank Salvador Rosauro-Alcaraz for providing the GLoBES~{.glb} file for the ESSnuSB experi\-ment. We would also like to thank Mattias Blennow, Marcos Dracos, Tord Ekel{\"o}f, and Enrique Fernandez-Martinez for valuable discussions. We acknowledge support by the project ESSnuSB (Contract No.~777419) under the Horizon 2020 Framework Programme funded by the European Union through the European Commission. This project is supported by the COST Action CA15139 ``Combining forces for a novel European facility for neutrino-antineutrino symmetry-violation discovery'' (EuroNuNet). In addition, T.O.~acknowledges support by the Swedish Research Council (Vetenskapsr{\aa}det) through Contract No.~2017-03934.
\end{acknowledgments}

\bibliography{ess_cp}

\begin{thebibliography}{24}
\expandafter\ifx\csname natexlab\endcsname\relax\def\natexlab#1{#1}\fi
\expandafter\ifx\csname bibnamefont\endcsname\relax
  \def\bibnamefont#1{#1}\fi
\expandafter\ifx\csname bibfnamefont\endcsname\relax
  \def\bibfnamefont#1{#1}\fi
\expandafter\ifx\csname citenamefont\endcsname\relax
  \def\citenamefont#1{#1}\fi
\expandafter\ifx\csname url\endcsname\relax
  \def\url#1{\texttt{#1}}\fi
\expandafter\ifx\csname urlprefix\endcsname\relax\def\urlprefix{URL }\fi
\providecommand{\bibinfo}[2]{#2}
\providecommand{\eprint}[2][]{\url{#2}}

\bibitem[{\citenamefont{Esteban et~al.}(2019)\citenamefont{Esteban,
  Gonzalez-Garcia, Hernandez-Cabezudo, Maltoni, and Schwetz}}]{Esteban:2018azc}
\bibinfo{author}{\bibfnamefont{I.}~\bibnamefont{Esteban}},
  \bibinfo{author}{\bibfnamefont{M.~C.} \bibnamefont{Gonzalez-Garcia}},
  \bibinfo{author}{\bibfnamefont{A.}~\bibnamefont{Hernandez-Cabezudo}},
  \bibinfo{author}{\bibfnamefont{M.}~\bibnamefont{Maltoni}}, \bibnamefont{and}
  \bibinfo{author}{\bibfnamefont{T.}~\bibnamefont{Schwetz}},
  \bibinfo{journal}{J. High Energy Phys.} \textbf{\bibinfo{volume}{01}},
  \bibinfo{pages}{106} (\bibinfo{year}{2019}), \eprint{1811.05487}.

\bibitem[{\citenamefont{Gariazzo et~al.}(2018)\citenamefont{Gariazzo,
  Archidiacono, de~Salas, Mena, Ternes, and T{\'o}rtola}}]{Gariazzo:2018pei}
\bibinfo{author}{\bibfnamefont{S.}~\bibnamefont{Gariazzo}},
  \bibinfo{author}{\bibfnamefont{M.}~\bibnamefont{Archidiacono}},
  \bibinfo{author}{\bibfnamefont{P.~F.} \bibnamefont{de~Salas}},
  \bibinfo{author}{\bibfnamefont{O.}~\bibnamefont{Mena}},
  \bibinfo{author}{\bibfnamefont{C.~A.} \bibnamefont{Ternes}},
  \bibnamefont{and}
  \bibinfo{author}{\bibfnamefont{M.}~\bibnamefont{T{\'o}rtola}},
  \bibinfo{journal}{J. Cosmol. Astropart. Phys.}
  \textbf{\bibinfo{volume}{1803}}, \bibinfo{pages}{011} (\bibinfo{year}{2018}),
  \eprint{1801.04946}.

\bibitem[{\citenamefont{Capozzi et~al.}(2018)\citenamefont{Capozzi, Lisi,
  Marrone, and Palazzo}}]{Capozzi:2018ubv}
\bibinfo{author}{\bibfnamefont{F.}~\bibnamefont{Capozzi}},
  \bibinfo{author}{\bibfnamefont{E.}~\bibnamefont{Lisi}},
  \bibinfo{author}{\bibfnamefont{A.}~\bibnamefont{Marrone}}, \bibnamefont{and}
  \bibinfo{author}{\bibfnamefont{A.}~\bibnamefont{Palazzo}},
  \bibinfo{journal}{Prog. Part. Nucl. Phys.} \textbf{\bibinfo{volume}{102}},
  \bibinfo{pages}{48} (\bibinfo{year}{2018}), \eprint{1804.09678}.

\bibitem[{\citenamefont{Baussan et~al.}(2014)}]{Baussan:2013zcy}
\bibinfo{author}{\bibfnamefont{E.}~\bibnamefont{Baussan}} \bibnamefont{et~al.}
  (\bibinfo{collaboration}{ESSnuSB Collaboration}), \bibinfo{journal}{Nucl.
  Phys. B} \textbf{\bibinfo{volume}{885}}, \bibinfo{pages}{127}
  (\bibinfo{year}{2014}), \eprint{1309.7022}.

\bibitem[{\citenamefont{Wildner et~al.}(2016)}]{Wildner:2015yaa}
\bibinfo{author}{\bibfnamefont{E.}~\bibnamefont{Wildner}} \bibnamefont{et~al.}
  (\bibinfo{collaboration}{ESSnuSB Collaboration}), \bibinfo{journal}{Adv. High
  Energy Phys.} \textbf{\bibinfo{volume}{2016}}, \bibinfo{pages}{8640493}
  (\bibinfo{year}{2016}), \eprint{1510.00493}.

\bibitem[{\citenamefont{Abe et~al.}(2018)}]{Abe:2016ero}
\bibinfo{author}{\bibfnamefont{K.}~\bibnamefont{Abe}} \bibnamefont{et~al.}
  (\bibinfo{collaboration}{Hyper-Kamiokande Collaboration}),
  \bibinfo{journal}{Prog. Theor. Exp. Phys.} \textbf{\bibinfo{volume}{2018}},
  \bibinfo{pages}{063C01} (\bibinfo{year}{2018}), \eprint{1611.06118}.

\bibitem[{\citenamefont{Prakash et~al.}(2012)\citenamefont{Prakash, Raut, and
  Sankar}}]{Prakash:2012az}
\bibinfo{author}{\bibfnamefont{S.}~\bibnamefont{Prakash}},
  \bibinfo{author}{\bibfnamefont{S.~K.} \bibnamefont{Raut}}, \bibnamefont{and}
  \bibinfo{author}{\bibfnamefont{S.~U.} \bibnamefont{Sankar}},
  \bibinfo{journal}{Phys. Rev. D} \textbf{\bibinfo{volume}{86}},
  \bibinfo{pages}{033012} (\bibinfo{year}{2012}), \eprint{1201.6485}.

\bibitem[{\citenamefont{Huber et~al.}(2002)\citenamefont{Huber, Lindner, and
  Winter}}]{Huber:2002mx}
\bibinfo{author}{\bibfnamefont{P.}~\bibnamefont{Huber}},
  \bibinfo{author}{\bibfnamefont{M.}~\bibnamefont{Lindner}}, \bibnamefont{and}
  \bibinfo{author}{\bibfnamefont{W.}~\bibnamefont{Winter}},
  \bibinfo{journal}{Nucl. Phys. B} \textbf{\bibinfo{volume}{645}},
  \bibinfo{pages}{3} (\bibinfo{year}{2002}), \eprint{hep-ph/0204352}.

\bibitem[{\citenamefont{Agarwalla et~al.}(2013)\citenamefont{Agarwalla,
  Prakash, and Sankar}}]{Agarwalla:2013ju}
\bibinfo{author}{\bibfnamefont{S.~K.} \bibnamefont{Agarwalla}},
  \bibinfo{author}{\bibfnamefont{S.}~\bibnamefont{Prakash}}, \bibnamefont{and}
  \bibinfo{author}{\bibfnamefont{S.~U.} \bibnamefont{Sankar}},
  \bibinfo{journal}{J. High Energy Phys.} \textbf{\bibinfo{volume}{07}},
  \bibinfo{pages}{131} (\bibinfo{year}{2013}), \eprint{1301.2574}.

\bibitem[{\citenamefont{Agarwalla et~al.}(2017)\citenamefont{Agarwalla, Ghosh,
  and Raut}}]{Agarwalla:2017nld}
\bibinfo{author}{\bibfnamefont{S.~K.} \bibnamefont{Agarwalla}},
  \bibinfo{author}{\bibfnamefont{M.}~\bibnamefont{Ghosh}}, \bibnamefont{and}
  \bibinfo{author}{\bibfnamefont{S.~K.} \bibnamefont{Raut}},
  \bibinfo{journal}{J. High Energy Phys.} \textbf{\bibinfo{volume}{05}},
  \bibinfo{pages}{115} (\bibinfo{year}{2017}), \eprint{1704.06116}.

\bibitem[{\citenamefont{Fukasawa et~al.}(2017)\citenamefont{Fukasawa, Ghosh,
  and Yasuda}}]{Fukasawa:2016yue}
\bibinfo{author}{\bibfnamefont{S.}~\bibnamefont{Fukasawa}},
  \bibinfo{author}{\bibfnamefont{M.}~\bibnamefont{Ghosh}}, \bibnamefont{and}
  \bibinfo{author}{\bibfnamefont{O.}~\bibnamefont{Yasuda}},
  \bibinfo{journal}{Nucl. Phys. B} \textbf{\bibinfo{volume}{918}},
  \bibinfo{pages}{337} (\bibinfo{year}{2017}), \eprint{1607.03758}.

\bibitem[{\citenamefont{Raut}(2017)}]{Raut:2017dbh}
\bibinfo{author}{\bibfnamefont{S.~K.} \bibnamefont{Raut}},
  \bibinfo{journal}{Phys. Rev. D} \textbf{\bibinfo{volume}{96}},
  \bibinfo{pages}{075029} (\bibinfo{year}{2017}), \eprint{1703.07136}.

\bibitem[{\citenamefont{Chatterjee et~al.}(2017)\citenamefont{Chatterjee,
  Pasquini, and Valle}}]{Chatterjee:2017xkb}
\bibinfo{author}{\bibfnamefont{S.~S.} \bibnamefont{Chatterjee}},
  \bibinfo{author}{\bibfnamefont{P.}~\bibnamefont{Pasquini}}, \bibnamefont{and}
  \bibinfo{author}{\bibfnamefont{J.~W.~F.} \bibnamefont{Valle}},
  \bibinfo{journal}{Phys. Lett. B} \textbf{\bibinfo{volume}{771}},
  \bibinfo{pages}{524} (\bibinfo{year}{2017}), \eprint{1702.03160}.

\bibitem[{\citenamefont{Blennow et~al.}(2015)\citenamefont{Blennow, Coloma, and
  Fernandez-Martinez}}]{Blennow:2014sja}
\bibinfo{author}{\bibfnamefont{M.}~\bibnamefont{Blennow}},
  \bibinfo{author}{\bibfnamefont{P.}~\bibnamefont{Coloma}}, \bibnamefont{and}
  \bibinfo{author}{\bibfnamefont{E.}~\bibnamefont{Fernandez-Martinez}},
  \bibinfo{journal}{J. High Energy Phys.} \textbf{\bibinfo{volume}{03}},
  \bibinfo{pages}{005} (\bibinfo{year}{2015}), \eprint{1407.3274}.

\bibitem[{\citenamefont{Agarwalla et~al.}(2014)\citenamefont{Agarwalla,
  Choubey, and Prakash}}]{Agarwalla:2014tpa}
\bibinfo{author}{\bibfnamefont{S.~K.} \bibnamefont{Agarwalla}},
  \bibinfo{author}{\bibfnamefont{S.}~\bibnamefont{Choubey}}, \bibnamefont{and}
  \bibinfo{author}{\bibfnamefont{S.}~\bibnamefont{Prakash}},
  \bibinfo{journal}{J. High Energy Phys.} \textbf{\bibinfo{volume}{12}},
  \bibinfo{pages}{020} (\bibinfo{year}{2014}), \eprint{1406.2219}.

\bibitem[{\citenamefont{Chakraborty et~al.}(2018)\citenamefont{Chakraborty,
  Deepthi, and Goswami}}]{Chakraborty:2017ccm}
\bibinfo{author}{\bibfnamefont{K.}~\bibnamefont{Chakraborty}},
  \bibinfo{author}{\bibfnamefont{K.~N.} \bibnamefont{Deepthi}},
  \bibnamefont{and} \bibinfo{author}{\bibfnamefont{S.}~\bibnamefont{Goswami}},
  \bibinfo{journal}{Nucl. Phys. B} \textbf{\bibinfo{volume}{937}},
  \bibinfo{pages}{303} (\bibinfo{year}{2018}), \eprint{1711.11107}.

\bibitem[{\citenamefont{Bernabéu and Segarra}(2018)}]{Bernabeu:2018use}
\bibinfo{author}{\bibfnamefont{J.}~\bibnamefont{Bernabéu}} \bibnamefont{and}
  \bibinfo{author}{\bibfnamefont{A.}~\bibnamefont{Segarra}},
  \bibinfo{journal}{J. High Energy Phys.} \textbf{\bibinfo{volume}{11}},
  \bibinfo{pages}{063} (\bibinfo{year}{2018}), \eprint{1807.11879}.

\bibitem[{\citenamefont{Huber et~al.}(2005)\citenamefont{Huber, Lindner, and
  Winter}}]{Huber:2004ka}
\bibinfo{author}{\bibfnamefont{P.}~\bibnamefont{Huber}},
  \bibinfo{author}{\bibfnamefont{M.}~\bibnamefont{Lindner}}, \bibnamefont{and}
  \bibinfo{author}{\bibfnamefont{W.}~\bibnamefont{Winter}},
  \bibinfo{journal}{Comput. Phys. Commun.} \textbf{\bibinfo{volume}{167}},
  \bibinfo{pages}{195} (\bibinfo{year}{2005}), \eprint{hep-ph/0407333}.

\bibitem[{\citenamefont{Blennow et~al.}(2018)\citenamefont{Blennow,
  Fernandez-Martinez, Ota, and Rosauro}}]{ess_deliverable}
\bibinfo{author}{\bibfnamefont{M.}~\bibnamefont{Blennow}},
  \bibinfo{author}{\bibfnamefont{E.}~\bibnamefont{Fernandez-Martinez}},
  \bibinfo{author}{\bibfnamefont{T.}~\bibnamefont{Ota}}, \bibnamefont{and}
  \bibinfo{author}{\bibfnamefont{S.}~\bibnamefont{Rosauro}}
  (\bibinfo{collaboration}{ESSnuSB Collaboration}) (\bibinfo{year}{2018}),
  \bibinfo{note}{essnusb.eu/DocDB/public/ShowDocument?docid=205}.

\bibitem[{\citenamefont{Coloma et~al.}(2013)\citenamefont{Coloma, Huber, Kopp,
  and Winter}}]{Coloma:2012ji}
\bibinfo{author}{\bibfnamefont{P.}~\bibnamefont{Coloma}},
  \bibinfo{author}{\bibfnamefont{P.}~\bibnamefont{Huber}},
  \bibinfo{author}{\bibfnamefont{J.}~\bibnamefont{Kopp}}, \bibnamefont{and}
  \bibinfo{author}{\bibfnamefont{W.}~\bibnamefont{Winter}},
  \bibinfo{journal}{Phys. Rev. D} \textbf{\bibinfo{volume}{87}},
  \bibinfo{pages}{033004} (\bibinfo{year}{2013}), \eprint{1209.5973}.

\bibitem[{\citenamefont{Minakata and Nunokawa}(2001)}]{Minakata:2001qm}
\bibinfo{author}{\bibfnamefont{H.}~\bibnamefont{Minakata}} \bibnamefont{and}
  \bibinfo{author}{\bibfnamefont{H.}~\bibnamefont{Nunokawa}},
  \bibinfo{journal}{J. High Energy Phys.} \textbf{\bibinfo{volume}{10}},
  \bibinfo{pages}{001} (\bibinfo{year}{2001}), \eprint{hep-ph/0108085}.

\bibitem[{\citenamefont{Barger et~al.}(2002)\citenamefont{Barger, Marfatia, and
  Whisnant}}]{Barger:2001yr}
\bibinfo{author}{\bibfnamefont{V.}~\bibnamefont{Barger}},
  \bibinfo{author}{\bibfnamefont{D.}~\bibnamefont{Marfatia}}, \bibnamefont{and}
  \bibinfo{author}{\bibfnamefont{K.}~\bibnamefont{Whisnant}},
  \bibinfo{journal}{Phys. Rev. D} \textbf{\bibinfo{volume}{65}},
  \bibinfo{pages}{073023} (\bibinfo{year}{2002}), \eprint{hep-ph/0112119}.

\bibitem[{\citenamefont{Ghosh et~al.}(2016)\citenamefont{Ghosh, Ghoshal,
  Goswami, Nath, and Raut}}]{Ghosh:2015ena}
\bibinfo{author}{\bibfnamefont{M.}~\bibnamefont{Ghosh}},
  \bibinfo{author}{\bibfnamefont{P.}~\bibnamefont{Ghoshal}},
  \bibinfo{author}{\bibfnamefont{S.}~\bibnamefont{Goswami}},
  \bibinfo{author}{\bibfnamefont{N.}~\bibnamefont{Nath}}, \bibnamefont{and}
  \bibinfo{author}{\bibfnamefont{S.~K.} \bibnamefont{Raut}},
  \bibinfo{journal}{Phys. Rev. D} \textbf{\bibinfo{volume}{93}},
  \bibinfo{pages}{013013} (\bibinfo{year}{2016}), \eprint{1504.06283}.

\bibitem[{\citenamefont{Chakraborty et~al.}(2019)\citenamefont{Chakraborty,
  Goswami, Gupta, and Thakore}}]{Chakraborty:2019jlv}
\bibinfo{author}{\bibfnamefont{K.}~\bibnamefont{Chakraborty}},
  \bibinfo{author}{\bibfnamefont{S.}~\bibnamefont{Goswami}},
  \bibinfo{author}{\bibfnamefont{C.}~\bibnamefont{Gupta}}, \bibnamefont{and}
  \bibinfo{author}{\bibfnamefont{T.}~\bibnamefont{Thakore}},
  \bibinfo{journal}{J. High Energy Phys.} \textbf{\bibinfo{volume}{05}},
  \bibinfo{pages}{137} (\bibinfo{year}{2019}), \eprint{1902.02963}.

\end{thebibliography}
\end{document}